\documentclass{sigchi}

\usepackage{balance}       
\usepackage{graphics}      
\usepackage[T1]{fontenc}   
\usepackage{txfonts}
\usepackage{mathptmx}
\usepackage[pdflang={en-US},pdftex]{hyperref}
\usepackage{color}
\usepackage{booktabs}
\usepackage{textcomp}

\usepackage{changes}

\usepackage{microtype}        
\usepackage{ccicons}          

\usepackage{todonotes}

\usepackage{subcaption}
\usepackage{gensymb}

\def\plaintitle{Forecasting User Attention During Everyday Mobile Interactions Using Device-Integrated and Wearable Sensors}

\def\emptyauthor{}
\def\plainkeywords{Egocentric Vision; Handheld Mobile Device; Attention Shifts; Mobile Eye Tracking; Attentive User Interfaces}

\makeatletter
\def\url@leostyle{%
  \@ifundefined{selectfont}{
    \def\UrlFont{\sf}
  }{
    \def\UrlFont{\small\bf\ttfamily}
  }}
\makeatother
\def\pprw{8.5in}
\def\pprh{11in}

\definecolor{linkColor}{RGB}{6,125,233}
\hypersetup{%
  pdftitle={\plaintitle},
  pdfauthor={\emptyauthor},
  pdfkeywords={\plainkeywords},
  pdfdisplaydoctitle=true, 
  bookmarksnumbered,
  pdfstartview={FitH},
  colorlinks,
  citecolor=black,
  filecolor=black,
  linkcolor=black,
  urlcolor=linkColor,
  breaklinks=true,
  hypertexnames=false
}

\begin{document}

\CopyrightYear{2018}
\setcopyright{acmlicensed}
\conferenceinfo{MobileHCI '18,}{September 3--6, 2018, Barcelona, Spain}
\isbn{978-1-4503-5898-9/18/09}\acmPrice{\$15.00}
\doi{https://doi.org/10.1145/3229434.3229439}

\title{\plaintitle}

\numberofauthors{2}
\author{%
  \alignauthor{Julian Steil\\
  \affaddr{Max Planck Institute for Informatics,\\ Saarland Informatics Campus, Germany}\\
    \email{jsteil@mpi-inf.mpg.de}}\\
  \alignauthor{Philipp M\"uller\\
  \affaddr{Max Planck Institute for Informatics,\\ Saarland Informatics Campus, Germany}\\
    \email{pmueller@mpi-inf.mpg.de}}\\
    \alignauthor{Yusuke Sugano\\
    \affaddr{Graduate School of Information Science and Technology, Osaka University, Japan}\\
    \email{sugano@ist.osaka-u.ac.jp}}\\
  \alignauthor{Andreas Bulling\\
  \affaddr{Max Planck Institute for Informatics,\\ Saarland Informatics Campus, Germany}\\
    \email{bulling@mpi-inf.mpg.de}}\\
}

\maketitle

\begin{abstract}
Visual attention is highly fragmented during mobile interactions, but the erratic nature of attention shifts currently limits attentive user interfaces to adapting after the fact, i.e.\ after shifts have already happened.
We instead study \textit{attention forecasting} -- the challenging task of predicting users' gaze behaviour (overt visual attention) in the near future.
We present a novel long-term dataset of everyday mobile phone interactions, continuously recorded from 20 participants engaged in common activities on a university campus over 4.5 hours each (more than 90 hours in total).
We propose a proof-of-concept method that uses device-integrated sensors and body-worn cameras to encode rich information on device usage and users' visual scene.
We demonstrate that our method can forecast bidirectional attention shifts
and predict whether the primary attentional focus is on the handheld mobile device.
We study the impact of different feature sets on performance and discuss the significant potential but also remaining challenges of forecasting user attention during mobile interactions.

\end{abstract}

\category{H.5.m.}{Information Interfaces and Presentation
  (e.g.\ HCI)}{Miscellaneous}


\section{Introduction}
\label{sec:introduction}
\begin{figure}[t!]
    \centering
    \includegraphics[width=1.0\columnwidth]{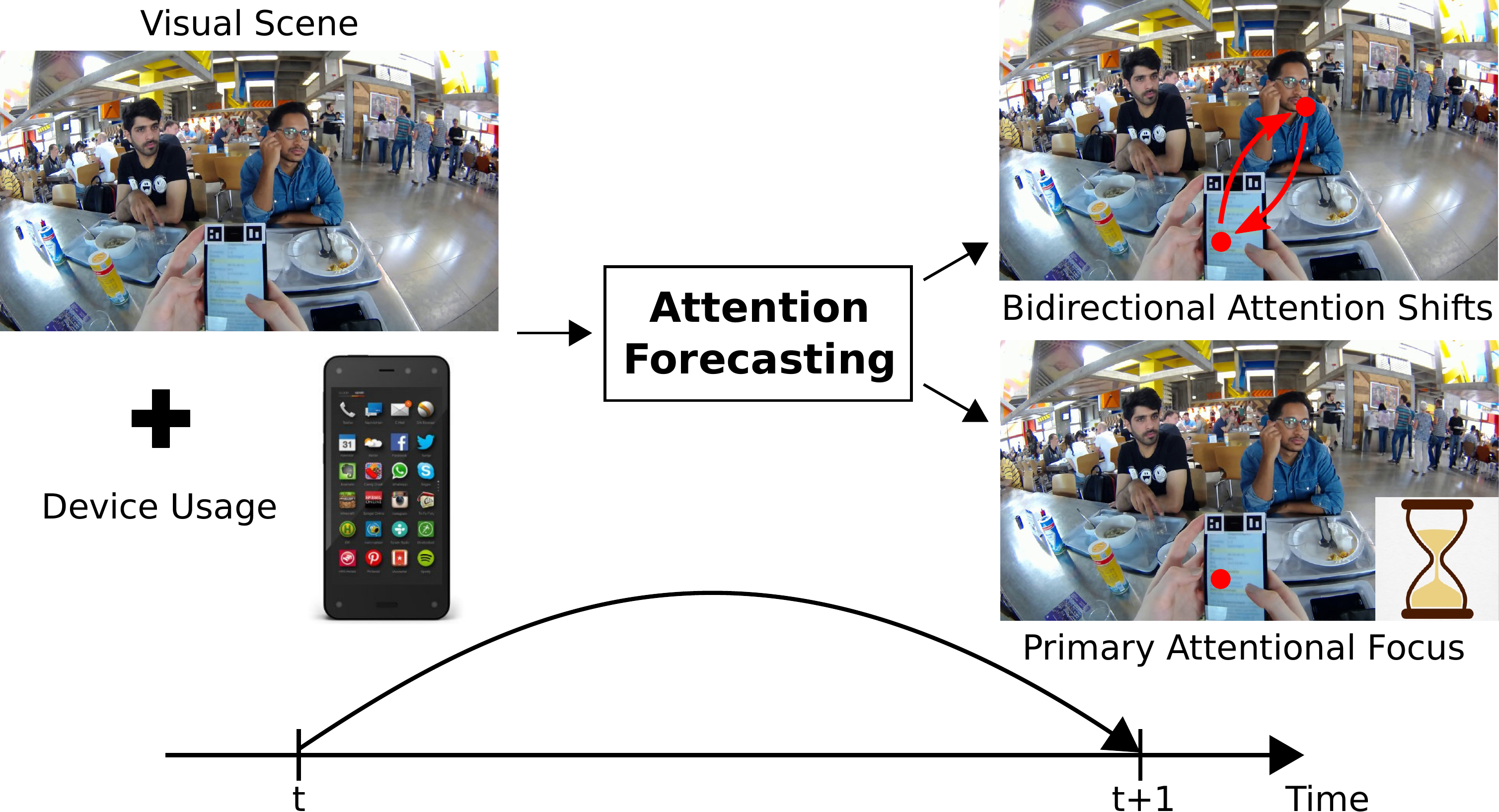}
    \caption{We propose a method to forecast temporal allocation of overt visual attention (gaze) during everyday interactions with a handheld mobile device. Our method uses information on users' visual scene as well as device usage to predict  attention shifts between mobile device and environment and primary attentional focus on the mobile device.}
    \label{fig:teaser}
\end{figure}

Sustained visual attention -- the ability to focus on a specific piece of information for a continuous amount of time without getting distracted -- has constantly diminished over the years~\cite{rubinstein2001executive}.
This trend is particularly prevalent for mobile interactions, during which user attention was shown to be highly fragmented~\cite{oulasvirta2005interaction}.
Active management of user attention has consequently emerged as a key research challenge in human-computer interaction~\cite{bulling16_computer}.
However, the capabilities of current mobile attentive user interfaces
are still severely limited.
Prior work mainly focused on estimating the point of gaze on the device screen using the integrated front-facing camera~\cite{holland2012eye,wood14_etra} or on using inertial sensors or application usage logs~\cite{choy2016looking,exler2016preliminary} to predict user engagement~\cite{mathur2016engagement,urh2016taskyapp} or boredom~\cite{pielot2015attention}.
In contrast, allocation of user attention across the device and environment has rarely been studied, and only using simulated sensors~\cite{miettinen2007predicting}.
Most importantly, existing attentive user interfaces are only capable to adapt \textit{after the fact}, i.e.\ after an attention shift has taken place~\cite{gutwin2017looking,kern2010gazemarks,mariakakis2015switchback}.

We envision a new generation of mobile attentive user interfaces that pro-actively adapt to imminent shifts of user attention, i.e.\ \textit{before} these shifts actually occur.
Pro-active adaptation promises exciting new applications.
For example, future attentive user interfaces could alert users in case of a (potentially dangerous) external event that they might miss due to predicted sustained attention to the mobile device.
Further, a predicted attention shift to the mobile device could trigger unlocking the device or loading the previous screen content to reduce interaction delays.
Finally, pro-active adaptations could also have significant impact in interruptibility research.
A future attentive user interface could show important information if user attention is predicted to continue to stay on the device or, inversely, alert users if an attention shift to the environment is predicted such that a mobile task cannot be finished in time, such as submitting a form or replying to a chat message.

The core requirement to realise such pro-active attentive user interfaces is their ability to predict users' \textit{future} allocation of overt visual attention during interactions with a mobile device.
We call this challenging new task \textit{attention forecasting}.
To facilitate algorithm development and evaluation for attention forecasting, we collected a multimodal dataset of 20 participants freely roaming a local university campus over several hours while interacting with a mobile phone.
Three annotators annotated the full dataset post-hoc with participants' current environment, indoor or outdoor location, their mode of locomotion, and whenever their attention shifted from the handheld device to the environment or back.
We then developed a computational method to forecast overt visual attention during everyday mobile interactions.
Our method uses device-integrated and head-worn IMU as well as computer vision algorithms for object class detection, face detection, semantic scene segmentation, and depth reconstruction.
We evaluate our method on the new dataset and demonstrate its effectiveness in predicting attention shifts between the mobile device and the environment as well as whether the primary attentional focus is on the device.

The specific contributions of this work are three-fold.
First, we propose \textit{attention forecasting} as the challenging new task of predicting future allocation of users' overt visual attention during everyday mobile interactions.
We propose a set of forecasting tasks that will facilitate pro-active adaptations to users' erratic attentive behaviour in future user interfaces.
Second, we present a novel 20-participant dataset of everyday mobile phone interactions.
The dataset including annotations will be made publicly available upon acceptance.
Third, we propose the first method to predict core characteristics of mobile attentive behaviour from device-integrated and wearable sensors.
We report a detailed evaluation of our method on the new dataset, and demonstrate the feasibility of predicting attention shifts between handheld mobile device and environment and the primary attentional focus on the device.

\section{Related Work}
\label{sec:related_work}

Our work is related to prior work on (1) user behaviour modelling and (2) gaze estimation on mobile devices as well as (3) computational modelling of egocentric attention.

\subsection{User Behaviour Modelling on Mobile Devices}

With the prevalence of sensor-rich mobile devices, modelling user behaviour, including gaze and attention, has gained significant popularity.
A large body of work investigated the use of device-integrated sensors to predict users' interruptibility~\cite{choy2016looking,exler2016preliminary,fogarty2005predicting,turner2015interruptibility,turner2017reachable}.
In particular, Obuchi et al.\ detected breaks in a user's physical activities using inertial sensors on the phone to push mobile notifications during these breaks~\cite{obuchi2016investigating}.
Dingler et al.\ used rapid serial visual presentation (RSVP) on a smartwatch in combination with eye tracking and detected when the reading flow was briefly interrupted, so that text presentation automatically paused or backtracked~\cite{dingler2016rsvp}.
Pielot et al.\ proposed a method to predict whether a participant will click on a notification and subsequently engage with the offered content~\cite{pielot2017beyond}.
Others aimed to predict closely related concepts, such as user engagement~\cite{mathur2016engagement,urh2016taskyapp}, boredom~\cite{pielot2015attention} or alertness~\cite{abdullah2016cognitive}.
Oulasvirta et al.\ investigated how different environments affected attention while users waited for a web page to load on a mobile phone~\cite{oulasvirta2005interaction}.
In a follow-up work, the same authors used a Wizard-of-Oz paradigm with simulated sensors to assess the feasibility of predicting time-sharing of attention,
including prediction of the number of glances, the duration of the longest glance, and the total and average durations of the glances to the mobile phone~\cite{miettinen2007predicting}.

Our work is the first to propose a method to predict attentive behaviour during everyday mobile interactions from real phone-integrated and body-worn sensors.
Another distinction from prior work is that our data collection constrained participants as little as possible, and specifically did not impose a scripted sequence of activities or environments.

\subsection{Gaze Estimation on Mobile Devices}

Estimating gaze on mobile devices has only recently started to receive increasing interest, driven by technical advances in gaze estimation and mobile eye tracking.
In an early work, Holland and Komogortsev proposed a learning-based method for gaze estimation on an unmodified tablet computer using the integrated front-facing camera~\cite{holland2012eye}.
More recently, Huang et al.\ presented a large-scale dataset and method for gaze estimation on tablets and conducted extensive evaluations on the impact of various factors on gaze estimation performance, such as ethnic background, glasses, or posture while holding the device~\cite{huang2015tabletgaze}.
Wood and Bulling used a model-based gaze estimation approach on an off-the-shelf tablet and achieved an average gaze estimation accuracy of 6.88\textdegree~at 12 frames per second~\cite{wood14_etra} while Vaitukaitis and Bulling combined methods from image processing, computer vision and pattern recognition to detect eye gestures using the built-in front-facing camera~\cite{vaitukaitis12_petmei}.
Jiang et al.\ proposed a method to estimate visual attention on objects of interest in the user's environment by jointly exploiting the phone's front- and rear-facing cameras~\cite{jiang2016vads} while Paletta et al.\ investigated accurate gaze estimation on mobile phones using a computer vision method to detect the phone in an eye tracker's scene video~\cite{paletta2014smartphone}.
While all of these works focused on estimating gaze spatially on the device screen, we are the first to predict attention allocation temporally.

\subsection{Computational Modelling of Egocentric Attention}

While bottom-up attention modelling, i.e.\ solely using image features, has been extensively studied in controlled laboratory settings, egocentric settings are characterised by a mix of bottom-up and top-down influences and are therefore less well explored.
Yamada et al.\ were among the first to predict egocentric attention using bottom-up image and egomotion information~\cite{yamada2011attention}.
Zhong et al.\ used a novel optical flow model to build a uniform spatio-temporal attention model for egocentric videos~\cite{zhong2016perception}.
Saliency models, which aim to predict which image regions most attract viewers' attention are an important type of computational model of visual attention~\cite{itti2000saliency}. 
However, none of these works aimed to predict attention during mobile interactions.
In addition, while we also use features extracted from egocentric video, we do not predict spatial attention distributions for the current video frame but use a short sequence of past frames (one second) to predict shifts of visual attention in the near future.

\begin{figure}[!t]
    \centering
    \includegraphics[width=\columnwidth]{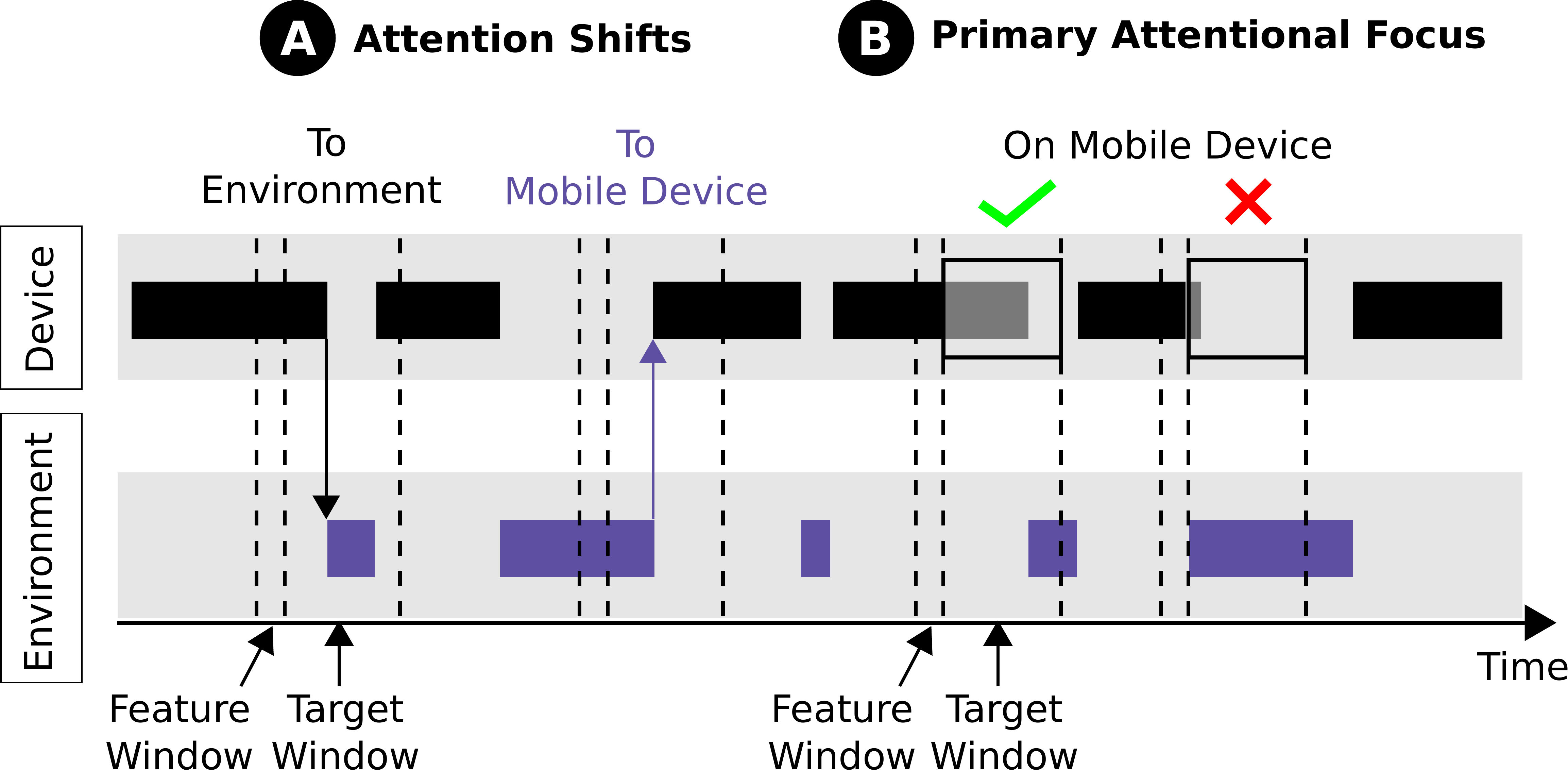}
   \caption{Overview of the different prediction tasks explored in this work: Prediction of attention shifts to the environment and (back) to the mobile device, and the primary attentional focus, i.e. whether attention is primarily on or off the device.
    }
    \label{fig:definitions}
\end{figure}

 \begin{figure*}[!t]
    \centering
    \includegraphics[width=1.0\textwidth]{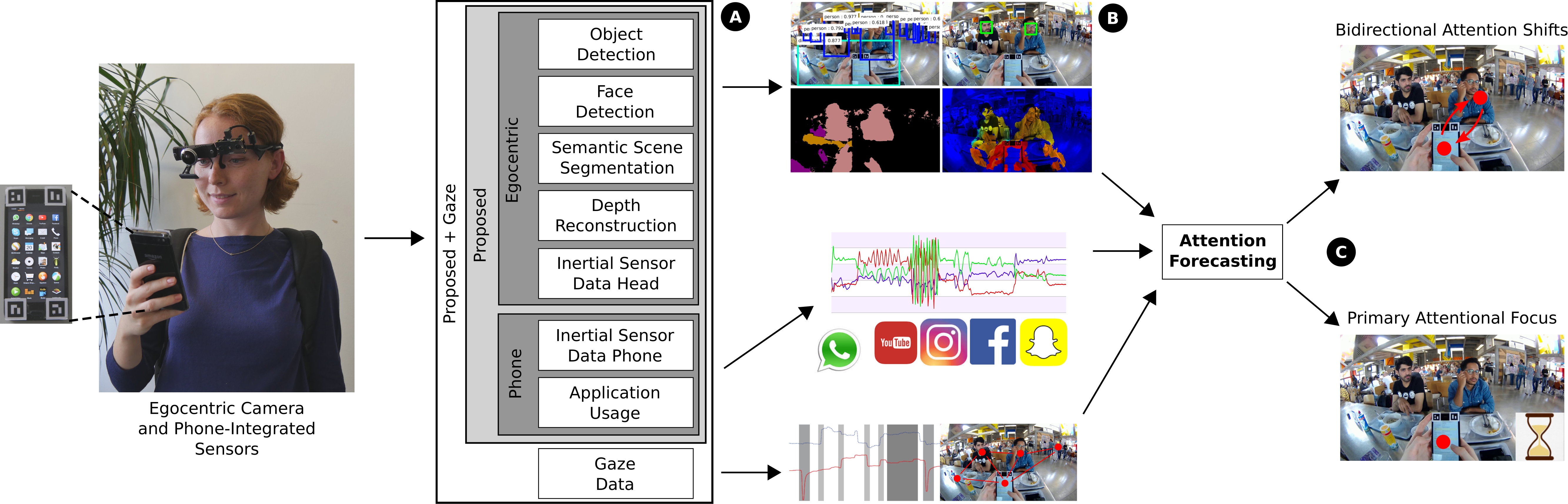}
    \caption{Overview of our method for attention forecasting during mobile interactions. Taking information on users' visual scene, mobile device (phone) and head inertial data, as well as on mobile app usage as input (A), our method extracts rich semantic information about the user's visual scene using state-of-the-art computer vision methods for object and face detection, semantic scene segmentation, and depth reconstruction (B). The method then extracts and temporally aggregates phone and visual features and takes eye tracking data into account to predict bidirectional attention shifts and the primary attentional focus on the phone (C). 
    }
    \label{fig:methods}
\end{figure*}

\section{Forecasting Mobile User Attention}
\label{sec:attention_prediction_using_egocentric_vision}

To be able to pro-actively adapt before users shift their attention,
attentive interfaces have to predict users' future attentive behaviour.
We call this new prediction task \textit{attention forecasting}.
Attention forecasting is similar in spirit to the tasks of user intention prediction as investigated, for example, in web search~\cite{cheng2010actively} or human-robot interaction~\cite{ravichandar2017human}, as well as player goal or plan recognition, studied in digital games~\cite{min2016player}.
In contrast to these lines of work, however, it specifically focuses on predicting fine-grained attentive behaviour and predictions at a moment-to-moment time scale.
Attention forecasting is already highly challenging in stationary desktop interaction settings given the significant variability and strong task dependence of users' attentive behaviour.
Forecasting users' attention is even more challenging during mobile interactions given the additional, as well as the large number of, potential visual attractors in the real-world environment.

In the following, we first propose a set of concrete prediction tasks within the attention forecasting paradigm and outline their potential use in future mobile attentive user interfaces.
A more extensive consideration of how attention forecasting could be used in the future can be found in the discussion section.
Afterwards, we propose a first proof-of-concept method that demonstrates the feasibility of predicting temporal attention allocation during everyday mobile interactions from real device-integrated and body-worn sensors.

\subsection{Prediction Tasks}

To guide future development of computational methods for attention forecasting during mobile interactions, we propose the following prediction tasks: prediction of \textit{Attention Shifts} to the environment and to the handheld mobile device, and \textit{Primary Attentional Focus} on the device.
\autoref{fig:definitions} illustrates these three prediction tasks for a sample attention allocation of a user.
During the segments marked in black the user's attention is on the mobile device, while during segments marked in purple the user's attention is in the environment.
In the following, we detail each of these prediction tasks.

\paragraph{Prediction of Attention Shifts}

The first prediction task deals with attention shifts from the mobile device to the environment, and from the environment back to the device (see \autoref{fig:definitions}A).
Attention shifts are a key characteristic of attentive behaviour and thus an important source of information for attentive user interfaces.
The task involves taking a certain time window for feature extraction, training a prediction model with this data, and using that model to predict whether an attention shift will happen during a subsequent target time window.
This task assumes the user interface to already have knowledge about whether a user's attention is currently on the handheld device or not.
Such knowledge can be obtained, for example, by using a method for mobile gaze estimation~\cite{wood14_etra}.
Prediction of attention shifts could be used in different ways by an attentive user interface.
Attention shift prediction could be used to pro-actively support users to reorient themselves on a mobile device to smoothly get back to their previous task.
Similar to Obuchi et al., who used phone data, predicted attention shifts could also be used as breakpoints for push notifications~\cite{obuchi2016investigating}. 
These could, for example, be shown shortly before or after an attention shift is predicted to take place.
Finally, attention shift prediction could be used to automatically turn the screen on again if a shift to the handheld device is predicted to occur in the near future.

\paragraph{Prediction of the Primary Attentional Focus}
The last task focuses on predicting
whether users' attention will be primarily on the mobile device or off the device for a particular time window in the future (see \autoref{fig:definitions}B).
Knowledge of the primary attentional focus for an upcoming time window can be useful for different applications.
For example, it could be used to highlight messages or to manage user attention in such a way that the interface needs to change content or style of presentation to keep users' attention beyond the considered time window to finish a task.

\subsection{Proposed Method}

To explore the feasibility of these prediction tasks, and to establish a baseline performance on each of them, we developed a first method for attention forecasting.
Previous work demonstrated that information available on a mobile device itself, such as inertial data, GPS location, or application usage, can be used to predict engagement or interruptibility.
It is therefore conceivable that such information may also be useful to predict attention shifts to the handheld mobile device.
In contrast, detecting shifts to the environment requires information on the user's current environment.
This suggests combining the mobile device with wearable sensors, in particular egocentric cameras worn on the user's head.
Egocentric cameras represent a rich source of visual information on the user's environment as demonstrated by the rapidly growing literature on egocentric vision~\cite{betancourt2015evolution}.
Combined with the fact that an ever-increasing number of egocentric cameras are used in daily life (e.g.\ sports cameras, cameras readily integrated in HMDs, life-logging cameras, etc.), this makes them a not only promising but also practical sensing modality for attention forecasting.

Figure~\ref{fig:methods} provides an overview of our method.
Inputs to our method are egocentric, mobile device (phone), and gaze data.
Our method extracts information from the egocentric scene and depth videos using computer vision algorithms for object and face detection, semantic scene segmentation labels, scene category, and reconstructed depth data as well as head motion.
In addition, our method extracts features from a mobile phone, including the history of application usage and accelerometer, gyroscope, and magnetometer measurements as well as past gaze.
Our method finally uses these features in a machine learning framework for attention forecasting, specifically attention shifts between the mobile phone and the environment as well as the primary attentional focus on the phone.

\subsection{Feature Extraction}

\definecolor{grayone}{rgb}{0.8235, 0.8235, 0.8235}
\definecolor{graytwo}{rgb}{0.5882, 0.5882, 0.5882}
\begin{table}[t]
\centering
\makebox[0pt][c]{\parbox{1.0\columnwidth}{%
\small
\renewcommand*{\arraystretch}{1.0}

\begin{minipage}[t]{0.5cm}

    \begin{flushleft}
        \vspace{-3.29cm}
        {\fcolorbox{black}{white}{\footnotesize\rotatebox{90}{\hspace{2.22cm}Proposed + Gaze\hspace{2.22cm}}}}
    \end{flushleft}

\end{minipage}
\begin{minipage}[t]{0.5cm}

    \begin{flushleft}
        \vspace{-3.29cm}
        {\fcolorbox{black}{grayone}{\footnotesize\rotatebox{90}{\hspace{2.3cm}Proposed\hspace{2.3cm}}}}
    \end{flushleft}

\end{minipage}
\begin{minipage}[t]{0.5cm}

    \begin{flushleft}
        \vspace{-3.29cm}
        {\fcolorbox{black}{graytwo}{\footnotesize\rotatebox{90}{\hspace{1.315cm}Egocentric\hspace{1.315cm}}}}
        {\fcolorbox{black}{graytwo}{\footnotesize\rotatebox{90}{\hspace{0.2085cm}\textcolor{graytwo}{y}Phone\textcolor{graytwo}{y}\hspace{0.2085cm}}}}
    \end{flushleft}

\end{minipage}
\begin{minipage}[t]{6.0cm}

\begin{tabular}{p{1cm}p{5.0cm}}
\toprule
\textbf{Sensor}    &   \textbf{Features}  \\
\hline
\textbf{RGB camera} &
number of detected faces and pixel counts of object classes like person, car, and monitor from the semantic segmentation, and binary occurrence indicator, numbers of detected instances of each object class from object detection, 1-hot encoded scene classes,
 mean, min, max, standard deviation and entropy of saliency and objectness of the scene images\\

\midrule

 \textbf{Depth camera} &  mean, min, max, standard deviation and entropy of the depth map from the stereo camera\\
 \midrule

\textbf{Head IMU}     & mean, min, max, standard deviation, norm and slope of accelerometer and gyroscope \\

\midrule

\textbf{Phone} & mean, min, max, standard deviation, norm and slope of accelerometer, gyroscope and orientation sensor values; 1/0 features indicating touch events, screen on/off, and activity of each of the installed applications\\

\midrule

\textbf{Gaze}    &
fixation positions (x, y); objectness, saliency and depth values at gaze position\\
\bottomrule
\end{tabular}

\end{minipage}
}}
\vspace{-0.15cm}
\caption{Overview of the different sensors and corresponding features explored in this work.}
\label{tab:features}
\end{table}

We extract features from the head-mounted egocentric RGB and depth cameras, head IMU, mobile device (phone), and past gaze data recorded using a head-mounted eye tracker (see Table~\ref{tab:features} for a complete list of features used in this work).
These features include numerical features, such as pixel counts of semantic segmentations, entropy of objectness maps, and mean depth map values, as well as binary encodings like occurrence of a touch event or whether an application on the handheld device is active.
We aggregate features over a window by computing the mean, maximum, minimum, standard deviation and slope for numerical features, and the mean and the slope for binary features.
Prior works on eye-based activity recognition demonstrated that gaze behaviour is characteristic for different activities~\cite{bulling11_pami,bulling13_chi,steil15_ubicomp}.
It is therefore conceivable that gaze features may help to improve the performance of our method for attention forecasting.
Specifically, we calculate mean, min, max, standard deviation, norm and slope of the gaze positions (x, y) as well as objectness, saliency and depth values at that position.
For evaluation purposes, and with potential future applications in mind, we group these features into four feature groups (cf.\ Figure~\ref{fig:methods} and~\autoref{tab:features}):
\textit{Egocentric} (including RGB, depth, and head inertial features), \textit{Phone} (including only phone features), \textit{Proposed} (all features from \textit{Egocentric} and \textit{Phone}), as well as \textit{Proposed + Gaze} (including fixation characteristics).

\paragraph{Egocentric}

This feature group covers the egocentric RGB and depth camera, as well as a head inertial sensor.
The depth and inertial sensors we used just for the sake of reliable feature extraction, although they can also be estimated from the egocentric camera itself~\cite{liu2015deep}.
As described above, we extract the most information from the egocentric scene video because scene information can include triggers which lead to changes of attentive behaviour.
We obtain a coarse description of the scene by applying the scene recognition method of Wang et al. ~\cite{wang2015places205} to the video frames.
This method utilises a convolutional neural network to extract scene descriptions like ``office'' or ``library''.
As objects are potential targets for capturing attention, we obtain a more fine-grained description of the scene by applying the semantic scene segmentation approach of Zheng et al.~\cite{zheng2015conditional}.
Semantic scene segmentation labels each pixel in a scene image as belonging to a certain object class or to background. 
To this end, their method combines a deep neural network with a probabilistic graphical model, trained to obtain pixel-wise segmentations of 20 different object classes including persons, monitors and cars.
By encoding the occurrence of objects and also counting the number of pixels belonging to each object class, we obtain information about which objects take up the largest portion of the camera's field of view.
Another important aspect of objects in a scene is the count of their instantiations.
For example, gazing upon a dining hall can lead to a large number of ``person'' pixels, as does standing directly in front of another person.
By simply counting the number of ``person'' pixels, these two cases cannot be distinguished.
Thus, we employ the object class detection method by Ren et al.~\cite{ren15fasterrcnn} to obtain an estimate of the count of instances for each object class. 
In addition to people detection, we hypothesised that faces can help in predicting attention shifts, as they are well known to strongly draw the attention of an observer~\cite{sato2015attentional} and their presence is also indicative of social situations~\cite{haxby2002human}, constituting a highly distracting factor in the scene.
To this end, we apply a face detection approach~\cite{dlib09} and count the number of detected faces in the scene image.
Moreover, we extracted depth information to obtain physical structure of the scene
and mapped the depth map to the scene video via camera calibration.
With the calculation of saliency and objectness maps, we collect ancillary knowledge about the scene complexity.
As head poses can serve as a useful prior for gaze estimation~\cite{valenti2012combining}, we additionally extract inertial features from the head-mounted camera.

\paragraph{Phone}

This feature group covers inertial data, which consists of accelerometer, gyroscope and orientation information, as well as phone usage data, which consists of single app usage information, and whether touch events took place or the screen is on or off.
For that purpose we installed additional applications on the phone which were running in the background to log the movement of the phone and the user's phone usage.

\section{Data Collection}
\label{sec:data_collection}

Given the lack of a suitable dataset for algorithm development and evaluation, we conducted our own data collection.
Our goal was to record natural attentive behaviour during everyday interactions with a mobile phone.
The authors of \cite{oulasvirta2005interaction} leveraged the -- at the time -- long page loading times during mobile web search to analyse shifts of attention.
We followed a similar approach but adapted the recording procedure in several important ways to increase the naturalness of participants' behaviour and, in turn, the realism of the prediction task.
First, as page loading times have significantly decreased over the last 10 years, we instead opted to engage participants in chat sessions during which they had to perform web search tasks as in~\cite{oulasvirta2005interaction} and then had to wait for the next chat message.

To counter side effects due to learning and anticipation, we varied the waiting time between chat messages and search tasks.
Second, we did not perform a fully scripted recording, i.e.\ participants were not asked to follow a fixed route or perform particular activities in certain locations in the city, they were not accompanied by an experimenter, and the recording was not limited to about one hour.
Instead, we observed participants passively over several hours while they interacted with the mobile phone during their normal activities on a university campus.
For our study we recruited twenty participants (six females), aged between 22 and 31 years, using university mailing lists and study board postings.
Participants were students with different backgrounds and subjects.
All had normal or corrected-to-normal vision.

\subsection{Apparatus}

The recording system consisted of a PUPIL head-mounted eye tracker~\cite{Kassner14_ubicomp} with an additional stereo camera, a mobile phone, and a recording laptop carried in a backpack (see Figure~\ref{fig:methods} left).
The eye tracker featured one eye camera with a resolution of 640$\times$480 pixels recording a video of the right eye from close proximity with 30 frames per second, and a scene camera with a resolution of 1280$\times$720 pixels recording at 24 frames per second.
The original lens of the scene camera was replaced with a fisheye lens with a 175$^\circ$ field of view.
The eye tracker was connected to the laptop via USB.
In addition, we mounted a DUO3D MLX stereo camera to the eye tracker headset.
The stereo camera recorded a depth video with a resolution of 752$\times$480 pixels at 30 frames per second as well as head movements using its integrated accelerometer and gyroscope.
Intrinsic parameters of the scene camera were calibrated beforehand using the fisheye distortion model from OpenCV.
The extrinsic parameters between the scene camera and the stereo camera were also calibrated.
The laptop ran the recording software and stored the timestamped egocentric, stereo, and eye videos.

Given the necessity to root the phone to record touch events and application usage, similar to~\cite{oulasvirta2005interaction} we opted to provide a mobile phone on which all necessary data collection software was pre-installed and validated to run robustly.
For participants to ``feel at home'' on the phone, we encouraged them to install any additional software they desired and to fully customise the phone to their needs prior to the recording.
Usage logs confirmed that participants indeed used a wide variety of applications, ranging from chat software, to the browser, mobile games, and maps.
To robustly detect the phone in the egocentric video and thus help with the ground-truth annotation, we attached visual markers to all four corners of the phone (see Figure~\ref{fig:methods} left).
We used WhatsApp to converse with the participants and to log accurate timestamps for these conversations~\cite{church2013s}.
Participants were free to save additional numbers from important contacts, but no one transferred their whole WhatsApp account to the study phone.
We used the Log Everything logging software to log phone inertial data and touch events~\cite{Weber:2014}, and the Trust Event Logger to log the current active application as well as whether the mobile phone screen was turned on or off.

\subsection{Procedure}

After arriving in the lab, participants were first informed about the purpose of the study and asked to sign a consent form.
We did not reveal which parts of the recording would be analysed later so as not to influence their behaviour.
Participants could then familiarise themselves with the recording system and customise the mobile phone, e.g. install their favourite apps, log in to social media platforms, etc.
Afterwards, we calibrated the eye tracker using the calibration procedure implemented in the PUPIL software~\cite{Kassner14_ubicomp}.
The calibration involved participants standing still and following a physical marker that was moved in front of them to cover their whole field of view.

To obtain some data from similar places on the university campus, we asked participants to visit three places at least once (a canteen, a library, and a caf\'e) and to not stay in any self-chosen place for more than 30 minutes.
Participants were further asked to stop the recording after about one and a half hours so we could change the laptop's battery pack and recalibrate the eye tracker.
Otherwise, participants were free to roam the campus, meet people, eat, or work as they normally would during a day at the university.
We encouraged them to log in to Facebook, check emails, play games, and use all pre-installed applications on the phone or install new ones.
Participants were also encouraged to use their own laptop, desktop computer, or music player if desired.

\begin{figure}[t]
    \centering
      \includegraphics[width=1.0\columnwidth]{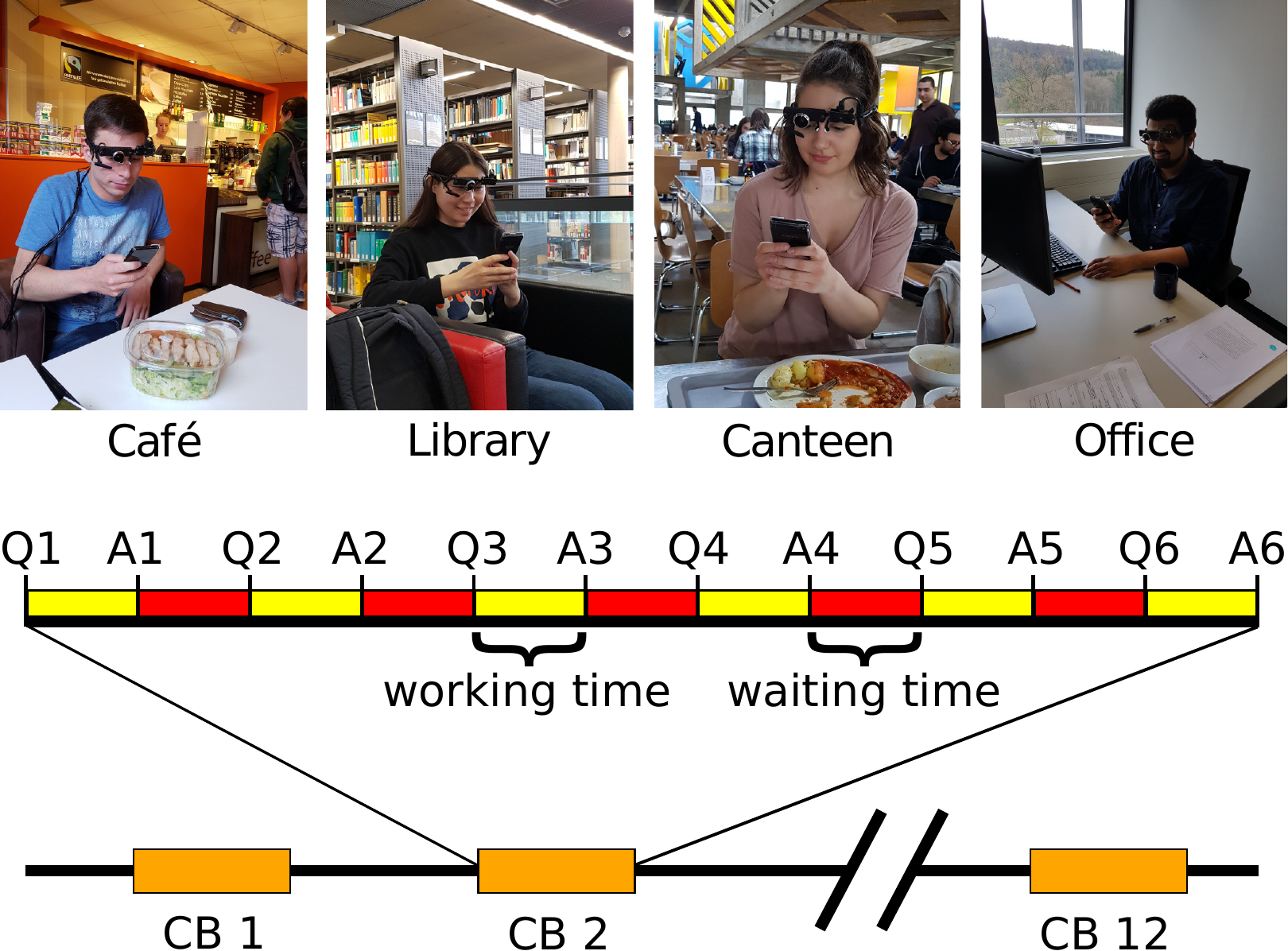}
    \caption{Participants were engaged in 12 chat blocks (CB) in different environments that were randomly distributed over their recording, which lasted in total about 4.5 hours. In each block, participants had to answer six questions, some of which required a short online search (Q1--Q6, working time), followed by waiting for the next question (waiting time). 
    }
    \label{fig:recordingplan}
\end{figure}

As illustrated in~\autoref{fig:recordingplan}, 12 chat blocks (CB) were distributed randomly over the whole recording.
Each block consisted of a conversation via WhatsApp during which the experimental assistant asked the participant six random questions (Q1--Q6) out of a pool of 72 questions.
Some questions could be answered with a quick online search, such as ``How many states are members of the European Union?'' or ``How long is the Golden Gate Bridge?''.
Similar to Oulasvirta et al.~\cite{oulasvirta2005interaction} we also asked simple demographic questions like ``What is the colour of your eyes?'' or ``What is your profession?'' that could be answered without an online search.
After each answer (A1--A6), participants had to wait for the next question.
This waiting time was varied randomly between 10, 15, 20, 30, and 45 seconds by the experimental assistant.
This was to avoid learning effects and to create a similar situation as in~\cite{oulasvirta2005interaction}.
This question-answering procedure was repeated until the sixth answer had been received, thus splitting each chat block into six working time segments (yellow) and five waiting time segments (red) (cf.\ Figure~\ref{fig:recordingplan}).
At the end of the recording, participants returned to the lab and completed a questionnaire about demographics and their mobile phone usage behaviour. In total, we recorded 1440 working and 1200 waiting segments over all participants. Statistics about our dataset are listed in Table~\ref{tab:dataset}.

\renewcommand*{\arraystretch}{1.15}
\begin{table}[b!]
\small
\begin{tabular}{p{4.5cm}p{0.8cm}p{0.8cm}p{0.8cm}}

\toprule
& \textbf{mean} & \textbf{std} & \textbf{total}\\

\midrule

\textbf{Working segments per question (sec)} & & &\\
Working time        &      40.29 & 11.27 & --:--\\
Time on mobile device      &      29.96 & 7.31 & --:--\\

\midrule

\textbf{Waiting segments per question (sec)} & & &\\
Waiting time        &      25.28 & 7.45 & --:--\\
Time on mobile device      &     11.02  & 4.26 & --:--\\

\midrule

\textbf{Attention shifts (quantity)} & & &\\
Shifts to environment & 248.85 & 107.22 & 4,957\\
Shifts to mobile device      & 259.90 & 106.88 & 5,178\\

\midrule

\textbf{Fixation time on/off screen (hh:mm)} & & &\\
 On        & 00:46 & 00:12 & 15:24\\
 Off       & 00:13 & 00:05 & 04:36\\

\midrule

\textbf{Environments (hh:mm)} & & &\\

Caf\'e      & 00:11 & 00:06 &  03:55\\
Corridor  & 00:12 & 00:12 &  04:08\\
Library   & 00:11 & 00:07 &     03:51\\
Canteen   & 00:08 & 00:06 &     02:50\\
Office    & 00:23 & 00:12 &     07:37\\
Street    & 00:04 & 00:06 &      01:20\\

\midrule

 \textbf{Indoor/Outdoor (hh:mm)}  & & &\\
Indoor      & 01:06 & 00:17 &      22:08\\
Outdoor     & 00:06 & 00:08 &     01:56\\

\midrule

\textbf{Modes of locomotion (hh:mm)} & & &\\
 Sit       & 01:02 & 00:14 &     20:49\\
 Stand     &  00:05 &  00:05 &     01:44\\
 Walk      & 00:04 & 00:04 &    01:31\\

\bottomrule
\end{tabular}

\caption{Statistics of the ground truth annotated chat block sequences with mean, standard deviation (std) and total time.}
    \label{tab:dataset}
\end{table}

\subsection{Data Preprocessing}

Fixations were detected from the raw gaze data using a dispersion-based algorithm with a duration threshold of 150ms and an angular threshold of $1\degree$~\cite{Kassner14_ubicomp}.
The 3D position of the mobile phone in the scene camera was estimated using visual markers (see Figure~\ref{fig:methods} left).
The position of the mobile phone surface was logged if at least two markers were visible in the scene camera. 
However, we only used the mobile phone detection as an aid for the ground truth annotation.

\subsection{Data Annotation}

Classifier training requires precise annotations of when an attention shift takes place and how long an attention span lasts.
Findlay and Gilchrist showed that in real-world settings, covert attention rarely deviates from the gaze location~\cite{findlay2003active}. 
Thus, we leveraged gaze as a reliable indicator of the user's current attentional focus.
Annotations were performed using videos extracted from the monocular egocentric video for the working/waiting time segments overlaid with gaze data provided by the eye tracker.
Three annotators were asked to annotate each chat block with information on participants' current environment (office, corridor, library, street, canteen, caf\'e), whether they were indoors or outdoors, their mode of locomotion (sitting, standing or walking), as well as when their attention shifted from the mobile device to the environment or back.

\section{Experiments}
\label{sec:experiments}

We conducted several experiments to evaluate the performance of our method for the different prediction tasks described before: attention shifts between the handheld mobile device and the environment and primary attentional focus on the device.
We evaluated our method for different time segments, i.e.\ while answering questions (\textit{working}) and while \textit{waiting} for the next question, as well as for the aforementioned four different feature groups.
For all experiments, we extracted features from a one-second window (feature window) and aimed to predict for a subsequent target window.
The choice of the one-second feature window was informed by preliminary experiments in which it showed superior performance compared to longer time windows.
For the target window size we investigated one, five, and ten seconds, reflecting that different applications might benefit from different time horizons when forecasting user attention.
Performance was calculated using the weighted F1 score.
The $F_1 \text{ score}=2*\frac{precision*recall}{precision+recall}$ is the harmonic mean of precision $\frac{TP}{TP+FP}$ and recall $\frac{TP}{TP+FN}$, where TP, FP, and FN represent frame-based true positive, false positive, and false negative counts, respectively.

We trained a random forest using the different features using a leave-one-person-out evaluation scheme, i.e., the data of n-1 participants was used for training, and of the last participant, for testing.
This procedure was repeated for all participants and the resulting F1 scores averaged over all iterations.
All hyperparameters (number of features, maximum depth and minimum samples at leaf nodes) were optimised via cross-validation on the training set.
We used a random subset of samples with a 50/50 distribution of positive and negative samples to avoid class imbalance.

\subsection{Performance for Different Prediction Tasks}

 \begin{figure}[!t]
    \centering
    \includegraphics[width=1.0\columnwidth]{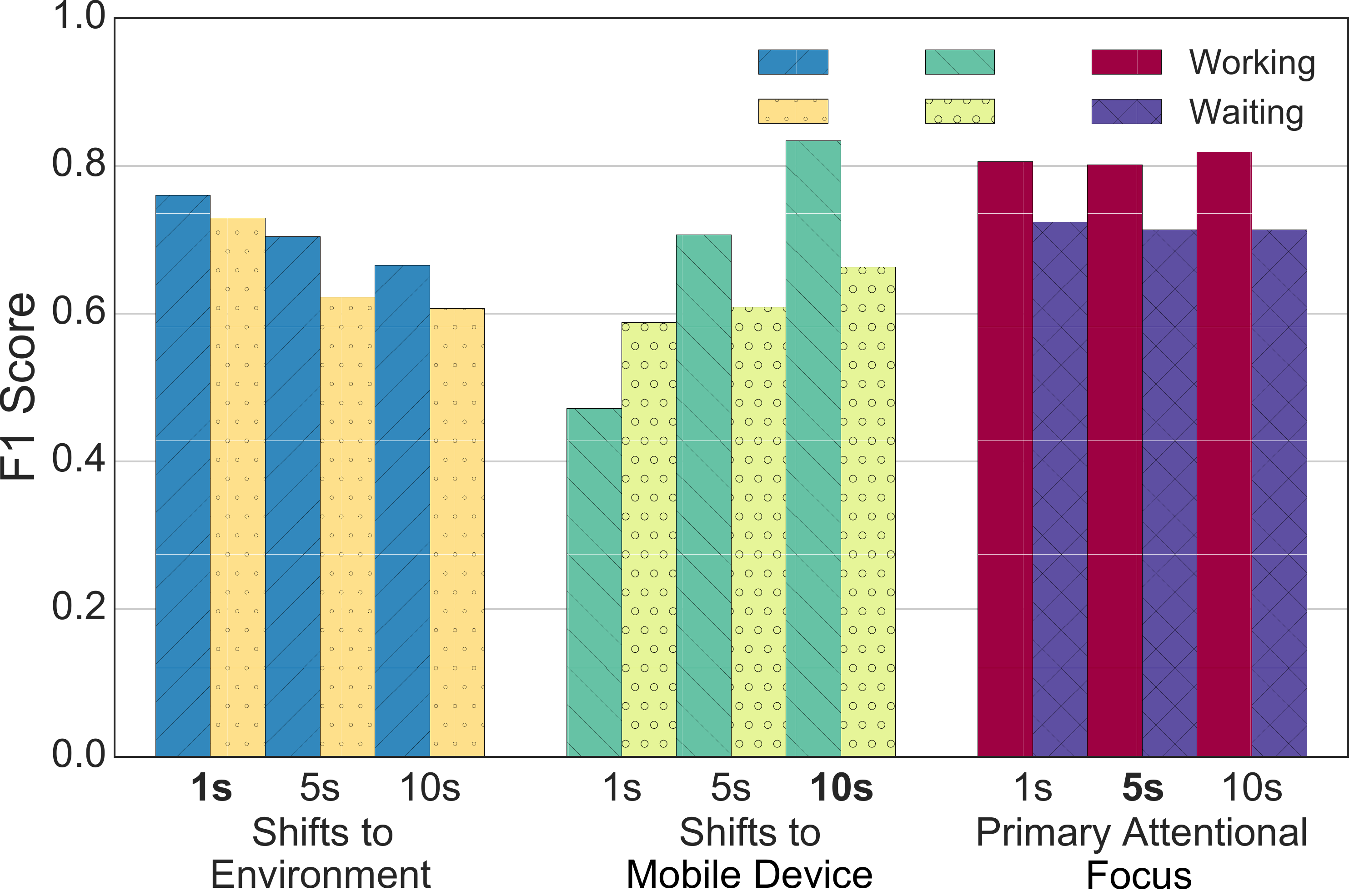}
    \caption{Performance analysis for shifts to environment, shifts to mobile device, and primary attentional focus for different target sizes (1s, 5s, 10s). 
    }
    \label{fig:performancecompare}
\end{figure}

\autoref{fig:performancecompare} summarises the performance of our proposed method for different target window sizes and the different prediction tasks.
As can be seen from the figure, the performance for predicting shifts to the environment decreases with increasing target window size, while for attention shifts to the mobile device an increase can be observed.
A possible interpretation for this is
that these shifts are often caused by distractors in the environment which result in a immediate reaction by the user.
When trying to predict shifts to the environment over a longer time interval in the future, such environmental distractors might not yet be present in the feature window.
To pro-actively pause interactions on a currently used device, a one-second target window for the prediction of shifts to the environment is sufficient, and it is not meaningful to choose a larger target window because the corresponding features do not contain the features necessary for a correct prediction.

On the other hand, a shift of attention back to the mobile device often lasts longer than just one second, as it might involve turning the head and picking up the mobile device, resulting in higher performance for longer target time intervals.
For the reduction of interaction delay when the attention shifts back to the device, a larger target window is needed anyway to restart the system or to load the previous screen content.
Moreover, predicted shifts to the mobile device can be used to avoid potential dangerous situations when the user shifts his/her attention to the device, e.g.\ when driving a car, an alert could warn the user to keep their attention on the street.
In such situations, predicting a shift to the device sufficiently early to still be able to intervene is required.
We therefore chose a target window size of ten seconds for shifts to the mobile device.

The primary attentional focus prediction is robust across target window size. Thus, longer target windows can be used to show notifications, or break long attention span prediction during dangerous situations.
We opted for a five-second target window for predicting the primary attentional focus.

\subsection{Prediction of Attention Shifts}
\begin{figure}[t]
\centering
  \includegraphics[width=\columnwidth]{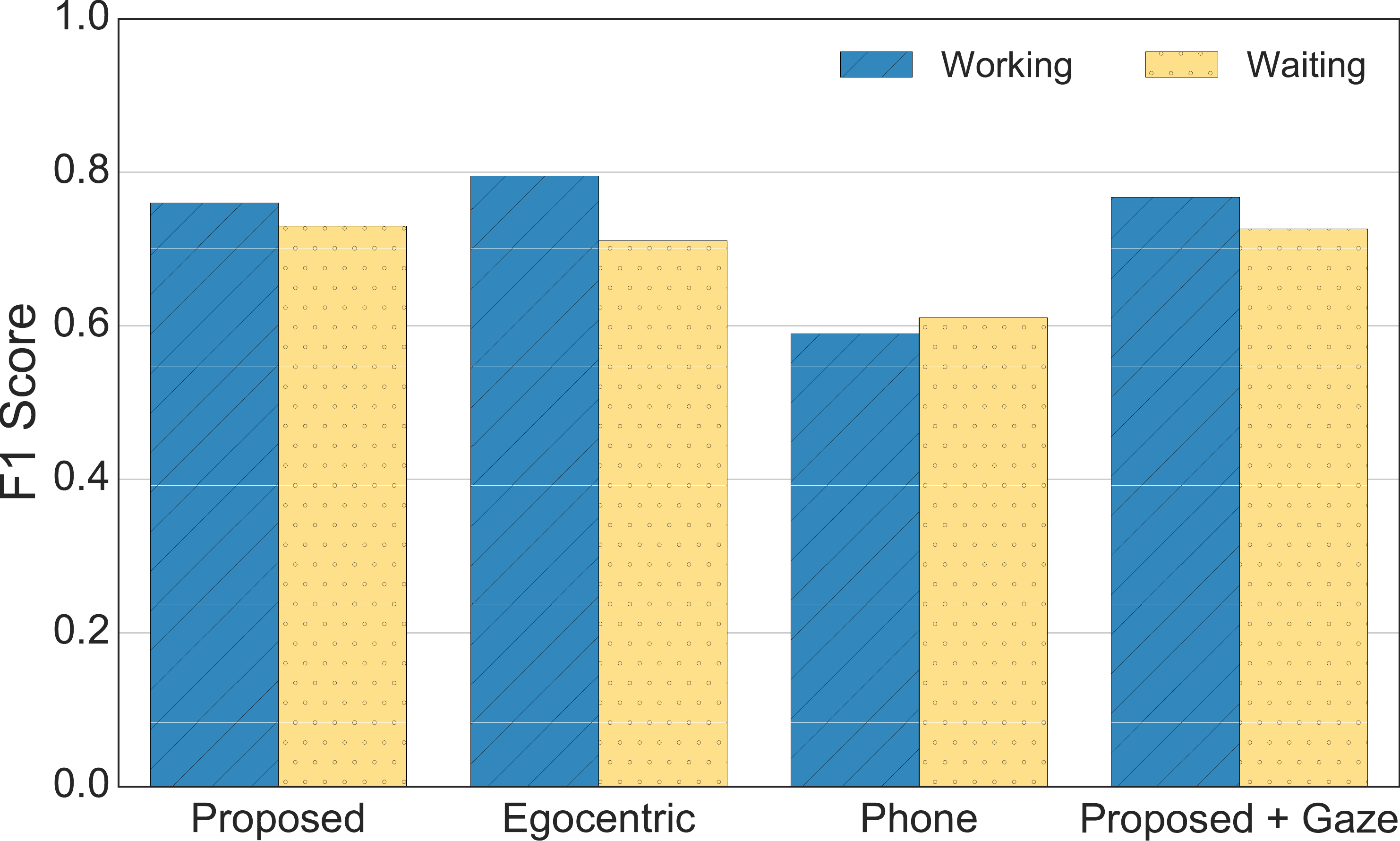}

      \vspace{0.1cm}

  \includegraphics[width=0.45\columnwidth]{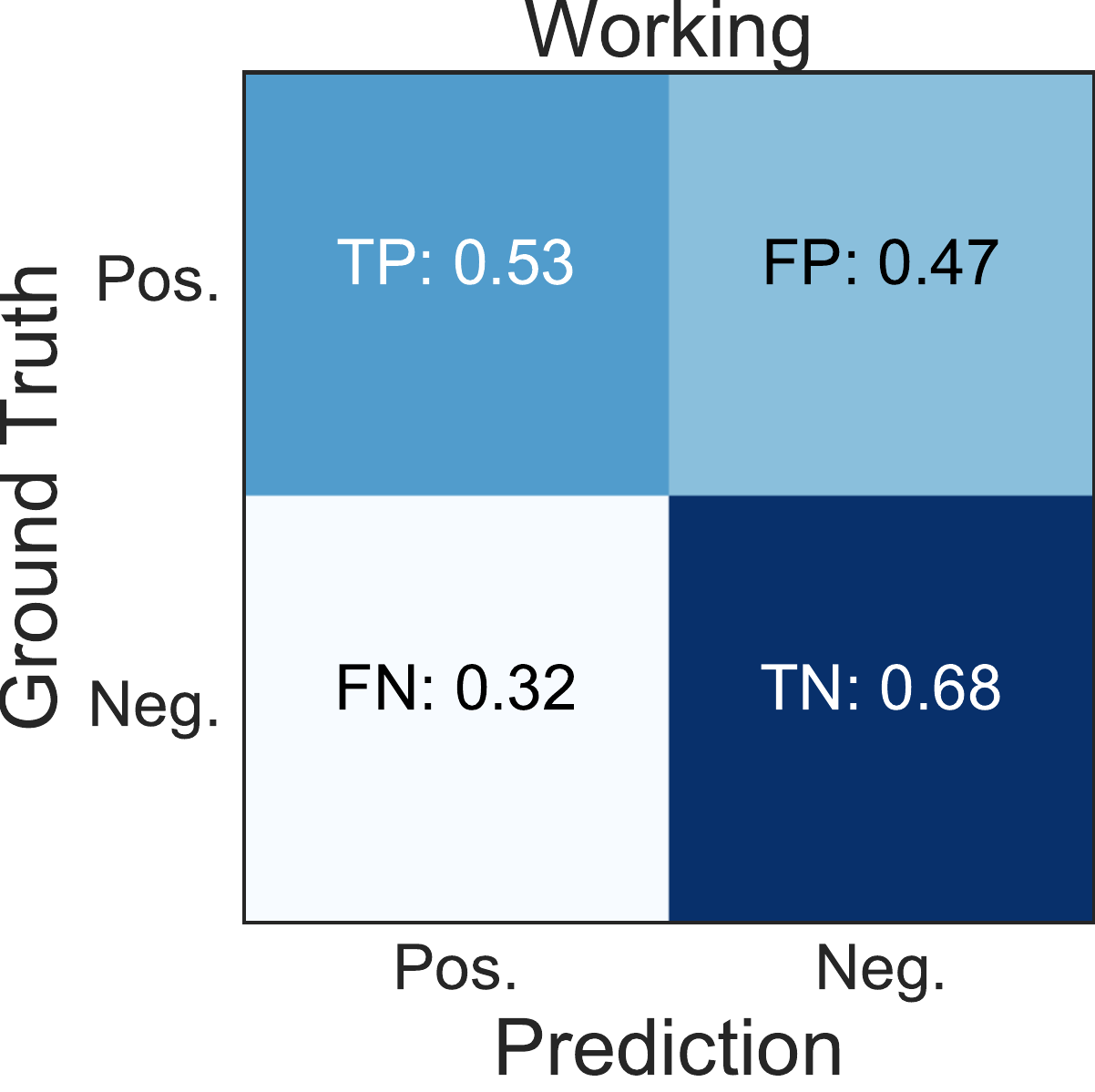} %
  \hspace{0.5cm}
  \includegraphics[width=0.45\columnwidth]{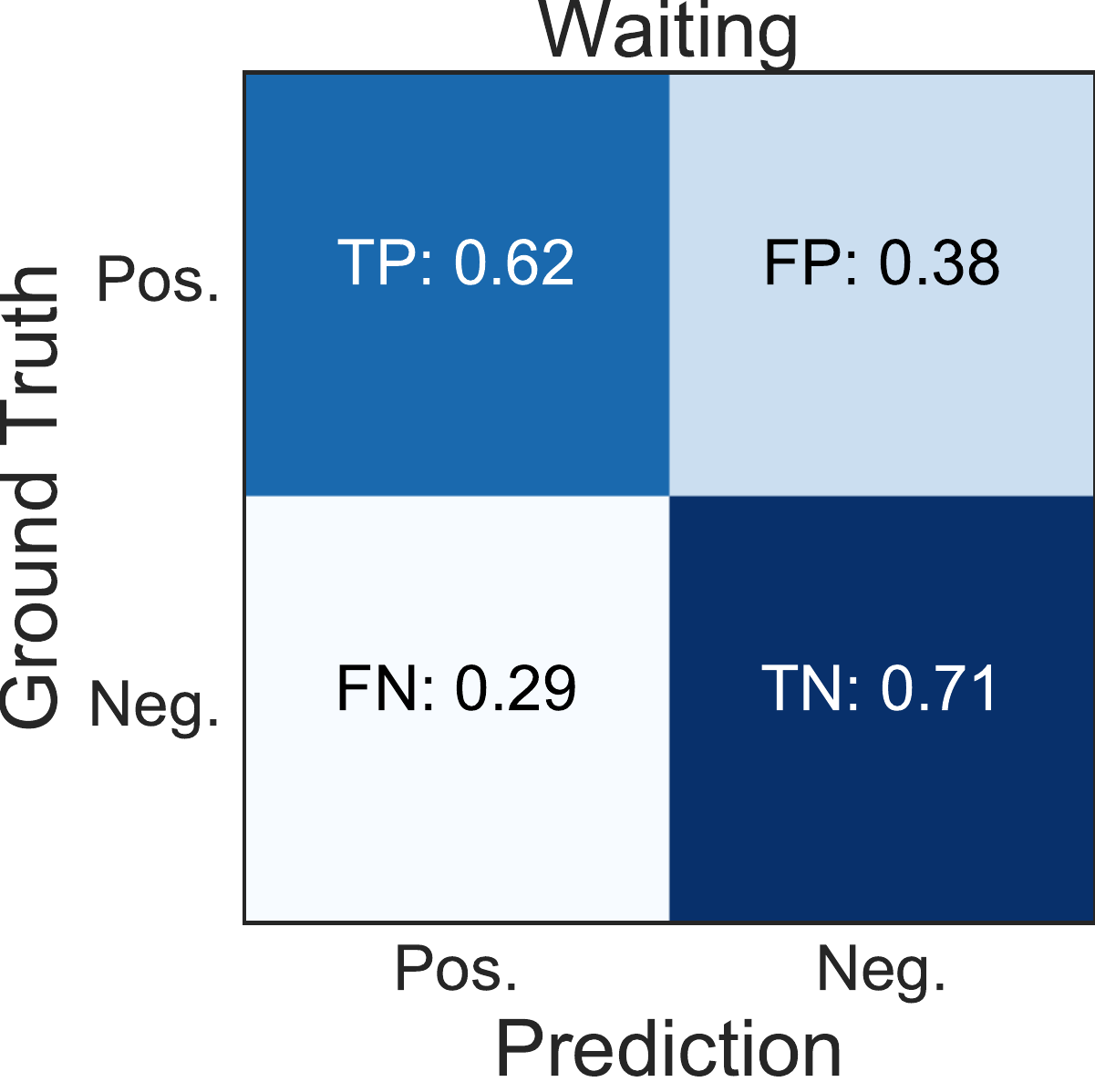}

  \vspace{-0.15cm}

\caption{Performance for predicting \textit{shifts to the environment} during working and waiting time segments for the different feature sets for a one-second target window, and confusion matrices for our proposed feature set.}
    \label{fig:stefeature}
\end{figure}

\begin{figure}[t]
\centering
  \includegraphics[width=\columnwidth]{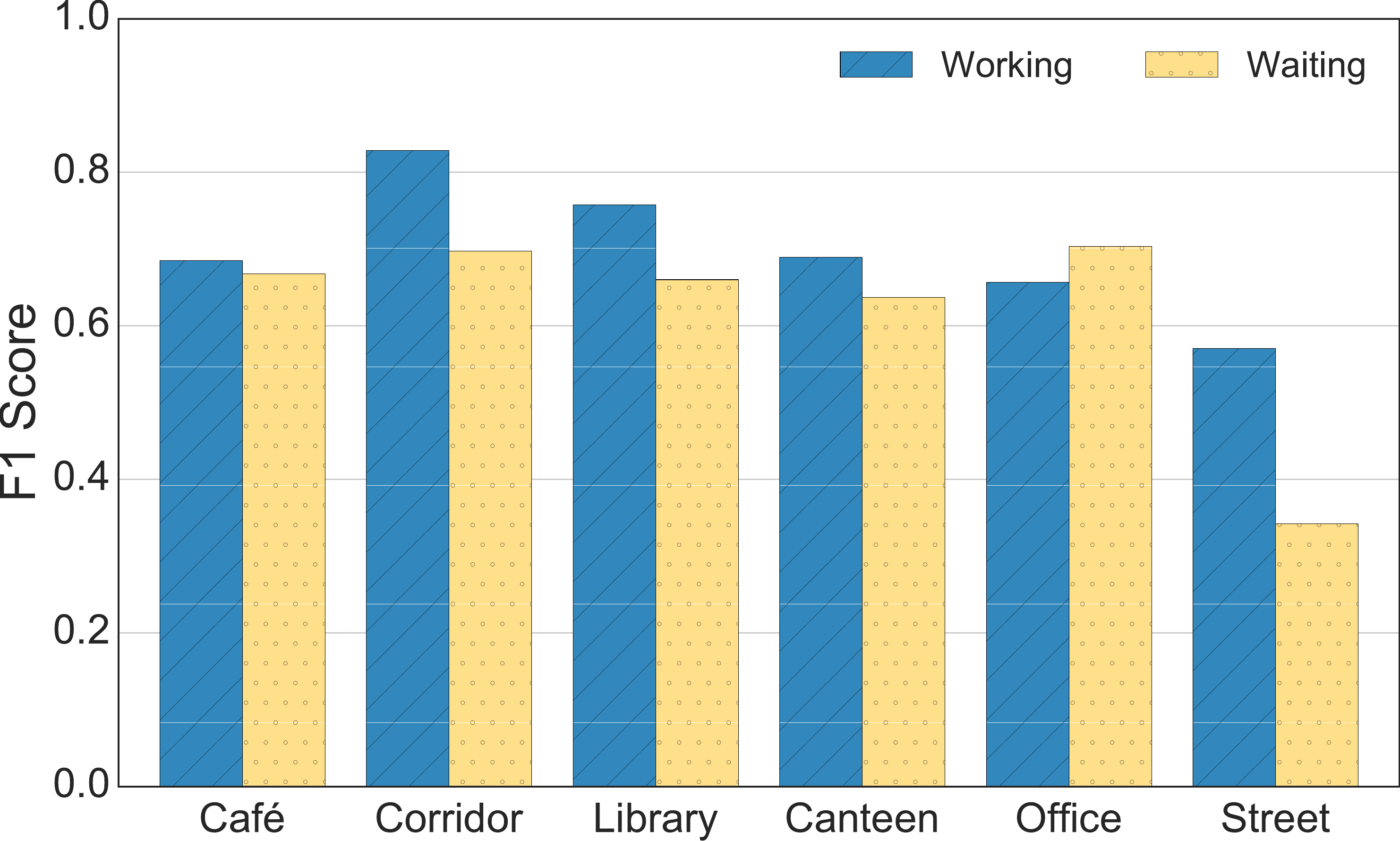}

\caption{Performance for predicting shifts to the environment for different real-world environments of our proposed feature set during working and waiting time segments.
    }
    \label{fig:environments}
 \end{figure}

\begin{figure}[t]
\centering
  \includegraphics[width=\columnwidth]{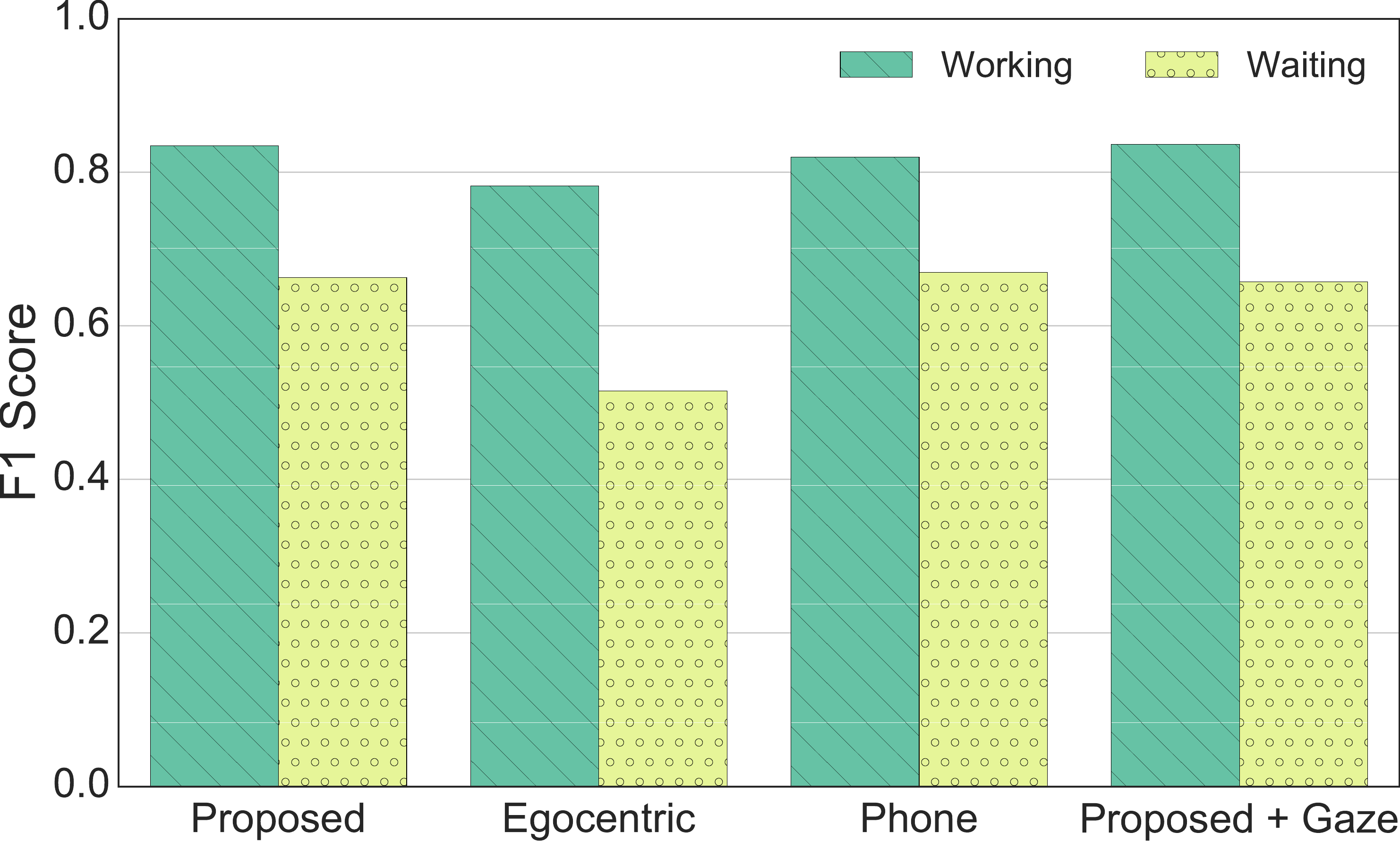}

      \vspace{0.1cm}

  \includegraphics[width=0.45\columnwidth]{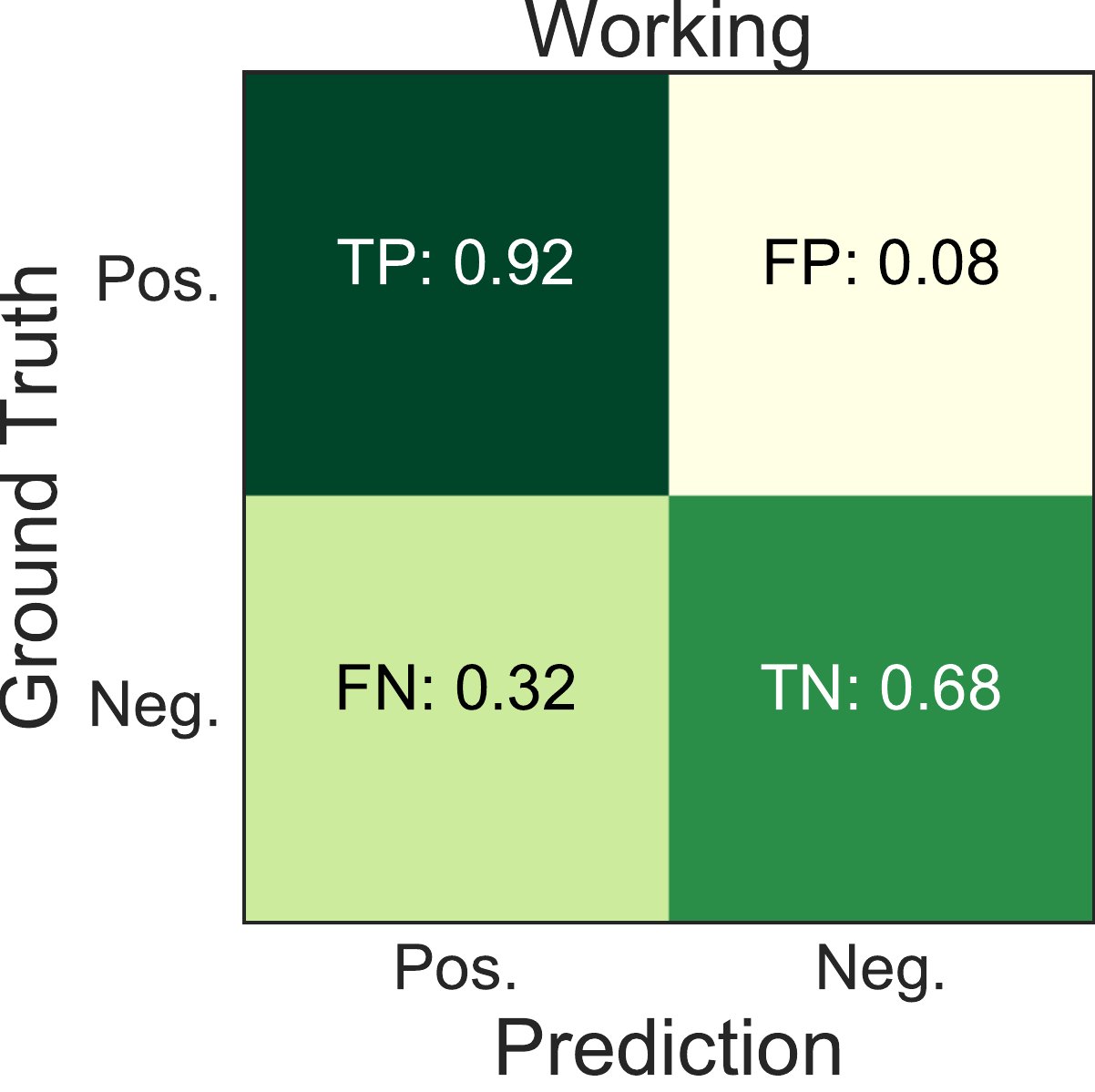} %
  \hspace{0.5cm}
  \includegraphics[width=0.45\columnwidth]{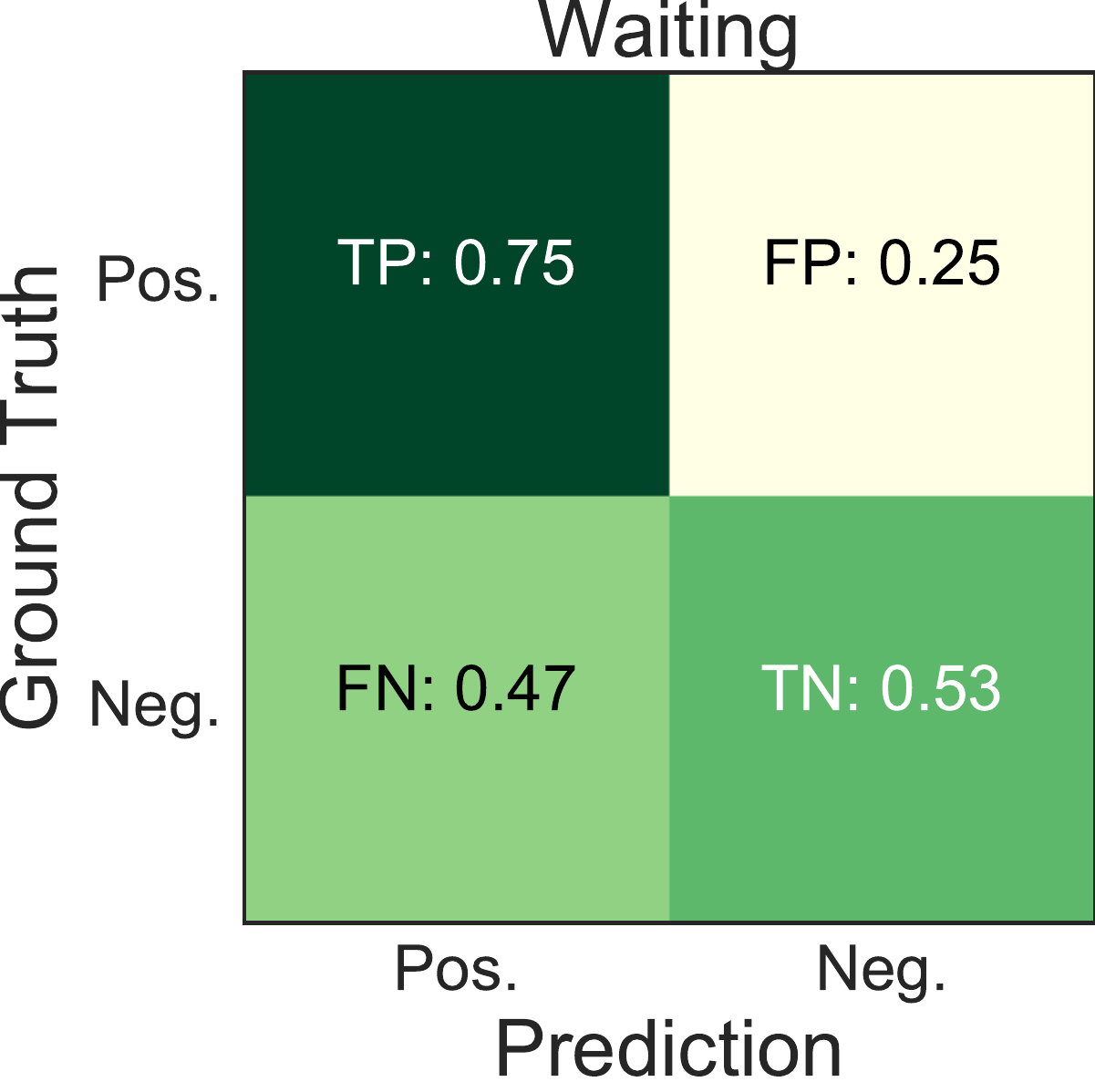}

  \vspace{-0.15cm}

\caption{Performance for predicting \textit{shifts to the mobile device} during working and waiting time segments for the different feature sets for a ten-second target window, and confusion matrices for our proposed feature set.}
    \label{fig:stsfeature}
\end{figure}

We first compared the performance of different feature sets for both attention shift prediction tasks.
Figure~\ref{fig:stefeature} shows the prediction performance of our method depending on feature sets used for both \textit{working} and \textit{waiting} time segments.
As can be seen from the figure, performance for predicting shifts to the environment is above chance level (F1 score 0.5) for all feature sets. 
This shows the effectiveness of our method for this challenging task.
However, we can see differences in the prediction performance between the working and waiting time segments and feature sets.
As expected, the \textit{Egocentric} sensor modality (F1 0.80)
performs competitively against the \textit{Proposed} feature combination (F1 0.76)
during working but also during waiting time segments.
During working segments performance is generally higher than during waiting segments except for the phone feature combination.
A possible explanation for this is that during working time, the task defines a certain phone interaction pattern (e.g. app usage, phone movement) with minor variability, whereas during waiting time the phone interaction can be chosen more freely (e.g. surfing the internet, using Facebook, playing games, chatting, etc.) and can induce different tendencies to switch one's attention to the environment.
A detailed feature analysis showed that especially during working time, detected faces from the scene camera are a helpful feature for the prediction of attention shifts to the environment.
The egocentric features, which are part of our proposed feature set, are the dominant ones for this task because shifts to the environment are mainly driven by attractors in our field of view.
However, having access to the smartphone state can also help the classifier.
The confusion matrices for predicting shifts to the environment show that the classifier achieves a good performance mainly on the negative training examples (i.e. no shift happening).

To further analyse the performance of our method for different environments, we evaluated our feature set in six environments each (see Figure~\ref{fig:environments}) during working and waiting time segments for the one-second target window. For the corridor and library environments our proposed feature set even exceeds an F1 score of 0.70, while the performance over all environments during working is higher than during waiting segments except for office environments. For the street environment, it is below 0.6 for working, and during waiting time segments even below 0.4, where participants are mainly focusing on the street and do not check their mobile devices as often as in the other environments.

For shifts to the mobile device the results are different from those for predicting shifts to the environment (see Figure~\ref{fig:stsfeature}).
With our proposed feature set we reach F1 scores of 0.66 during waiting and F1 scores of 0.83 during working time segments for the ten-second target window, respectively.
The competitive performance of phone features for the attention shift forecasting is caused by participants' natural device usage behaviour, which is characterised by picking up and moving the device or turning on its screen. 
Participants often held their phones in their hands out of the view of the camera, so there was a movement of the device followed by the shift to the device and a touch sequence to unlock the phone.
A detailed feature analysis confirmed that both actions were registered by the phone sensors and logging apps with F1 scores higher than 0.8 (phone IMU and application usage).
Features from the egocentric camera only resulted in chance-level performance, which indicates that the visual environment of the participant does not play a role in determining whether the attention will go back to the screen.
This is in line with our reasoning given above, indicating that poorly observable top-down factors influence shifts to the phone, as compared to better observable properties of the visual environment that might capture attention in a way that is more influenced by bottom-up processes.
In contrast to the prediction of shifts to the environment, the most errors occur for the negative examples, as indicated by the confusion matrices.

\subsection{Prediction of the Primary Attentional Focus}

\begin{figure}[t]
\centering
  \includegraphics[width=\columnwidth]{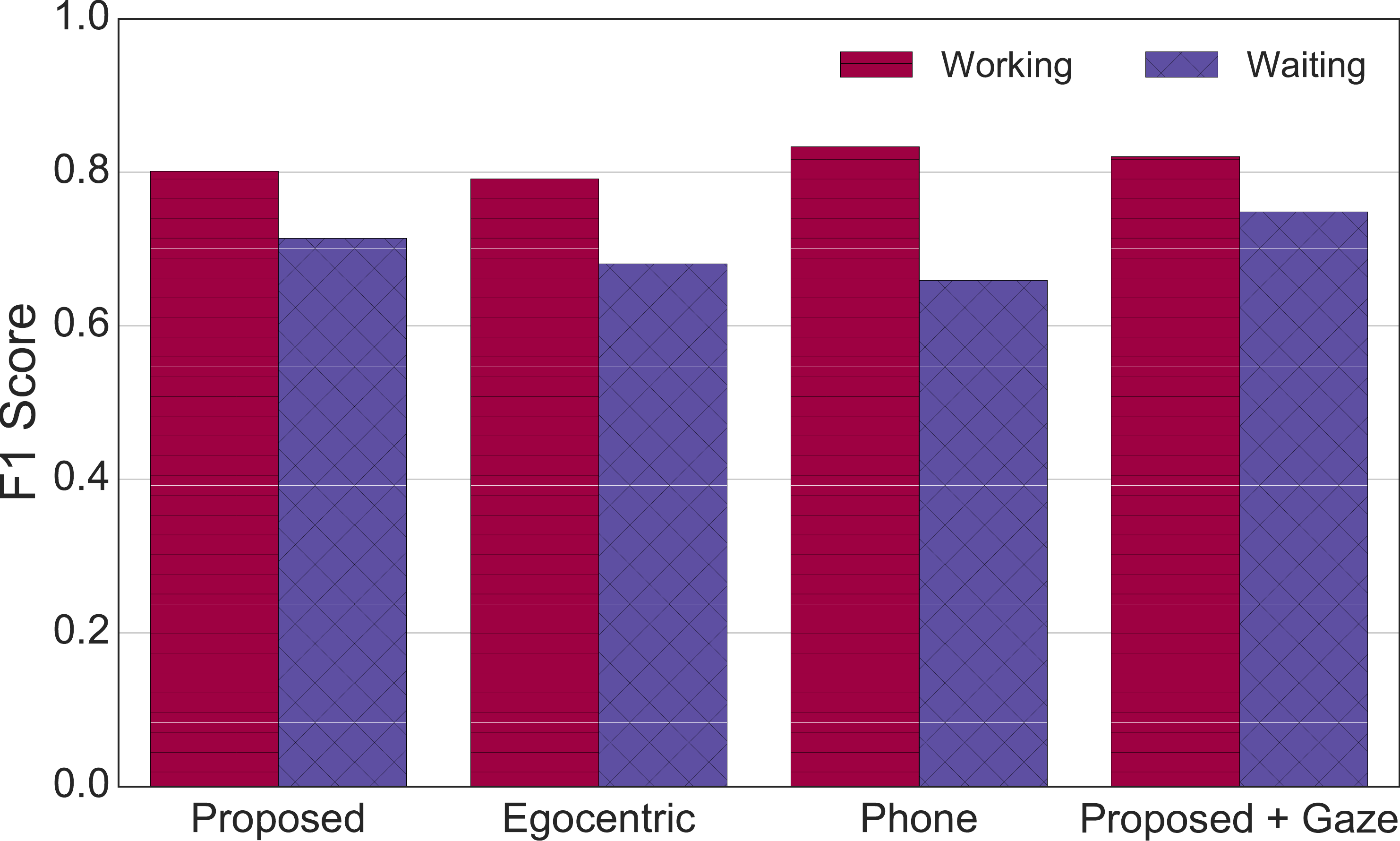}

      \vspace{0.1cm}

  \includegraphics[width=0.45\columnwidth]{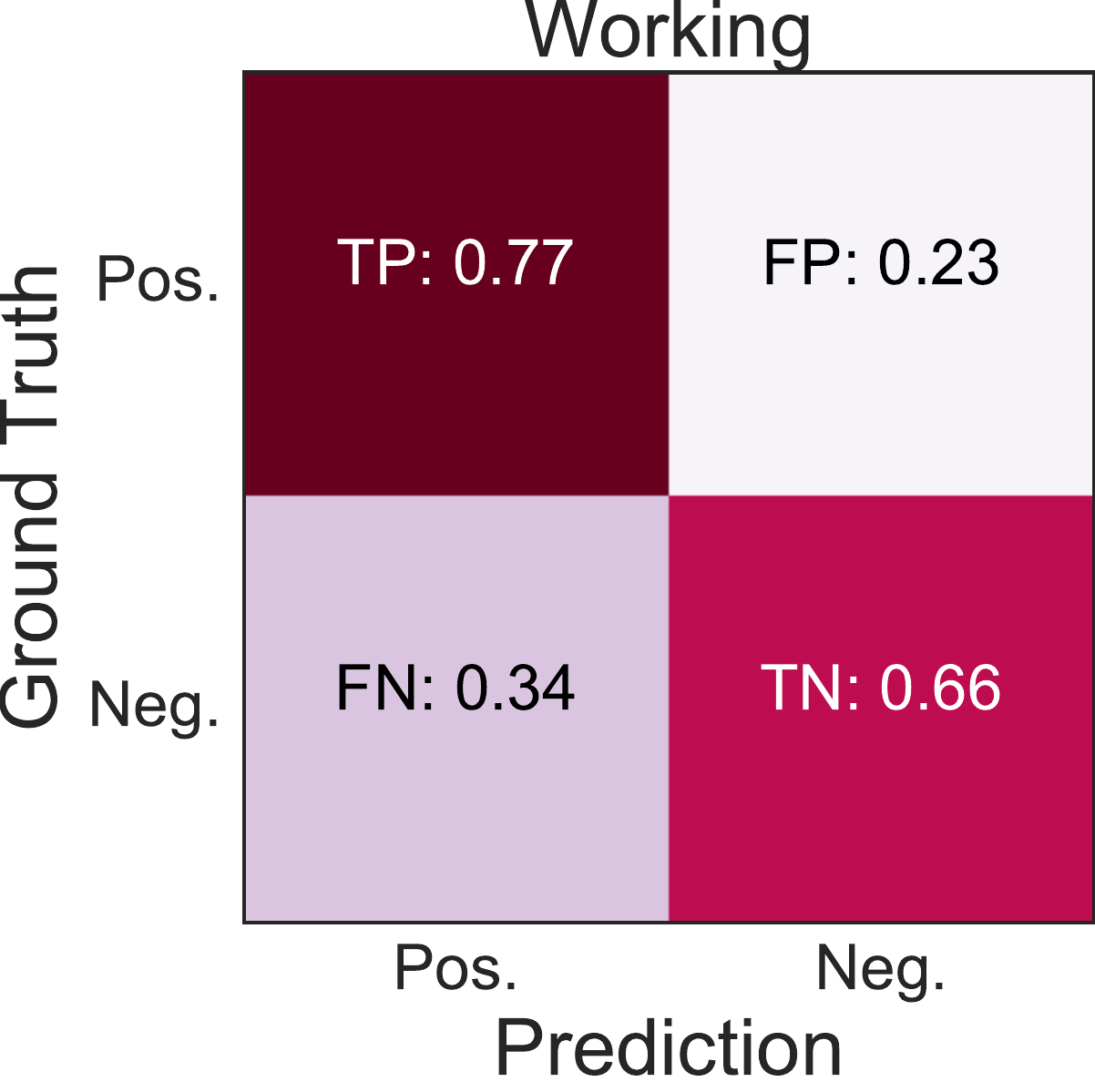} %
  \hspace{0.5cm}
  \includegraphics[width=0.45\columnwidth]{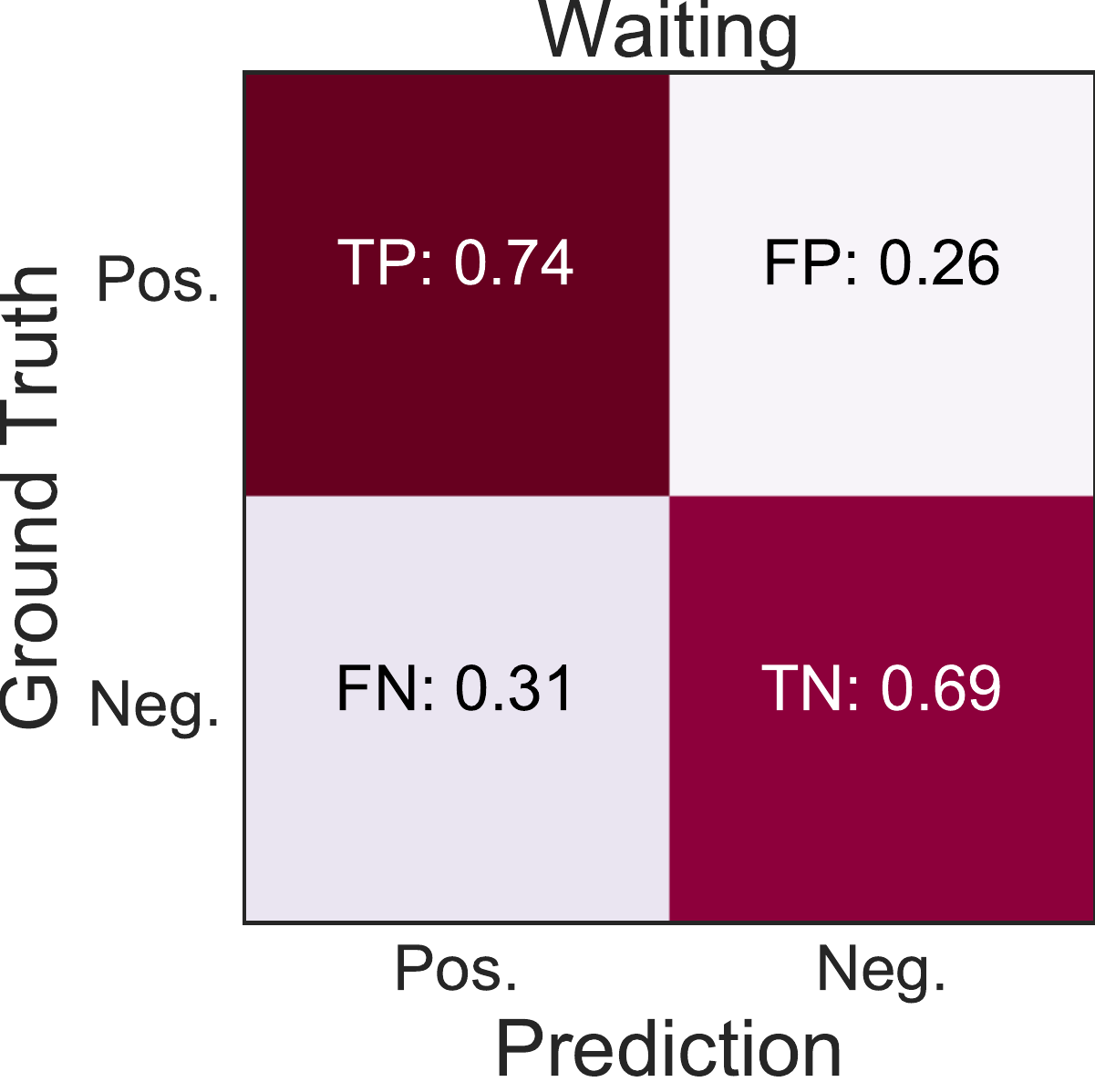}

  \vspace{-0.15cm}

\caption{Performance for \textit{primary attentional focus} on mobile device during working and waiting time segments for the different feature sets for a five-second target window, and confusion matrices for our proposed feature set.}
    \label{fig:aodfeature}

\end{figure}

Finally, we analysed the performance of our method for predicting the primary attentional focus on the mobile device.
As can be seen from Figure~\ref{fig:aodfeature}, for this prediction task, our method reaches an F1 score of more than 0.7 for both 
working and waiting time segments.
It can also be seen that combining features is helpful in all cases.
A detailed feature analysis
shows that head IMU, depth, and face features from the egocentric feature subsets, as well as the phone IMU, and app usage features, contribute especially to the good performance of our method.
Phone features show performance competitive to our proposed features during working but a lower performance during waiting time segments.
From a detailed feature analysis it can be seen that users' app usage patterns on the mobile device contributed especially to the performance.
The proposed feature combination can even be improved when taking gaze information into account, reaching an F1 performance larger than 0.8 during working and 0.75 during waiting time segments.
Thus, for this kind of prediction task, a full eye tracking system is a meaningful setup.
The increasing availability of mobile eye tracking as well as gaze estimation using the cameras readily integrated into laptop, tablets, and public displays~\cite{sugano2016aggregaze,wood14_etra,zhang15_cvpr,zhang18_pami} makes gaze another interesting source of information on users' future attentive behaviour.
The corresponding confusion matrices show that our approach performs clearly above chance on all ground truth classes.

\section{Discussion}
\label{sec:discussion}

The experiments demonstrated that our method can predict several key aspects of attentive behaviour during everyday mobile interactions using a combination of egocentric and device-integrated sensors.
Specifically, we showed that we can predict shifts between the handheld mobile device and environment, as well as the primary attentional focus, above chance level.
These results are promising for future mobile attentive user interfaces, particularly given the large variability in natural user behaviour and the large number of possible visual attractors in users' environments, and thus the difficulty of these prediction tasks.

\paragraph{Importance of Different Features}
For predicting shifts to the environment, egocentric features contributed most to the performance
(see Figure~\ref{fig:stefeature}).
A detailed feature analysis showed that face features especially, but also head IMU, semantic scene and depth features, contributed positively.
In contrast, phone features showed the best performance for predicting attention shifts back to the mobile device (see Figure~\ref{fig:stsfeature}).
The chance-level performance for the egocentric features suggested that shifts to the mobile device were less influenced by the environment, especially during waiting time segments.
This was to be expected given that such shifts are typically triggered by events on the mobile device, such as an incoming chat message or notification.

Our method performed robustly for predicting attention shifts in different environments, with performance peaking for working and waiting time segments in the corridor (see~\autoref{fig:environments}).
Results for predicting the primary attentional focus (a binary classification task) suggested that information readily available on the handheld device is most informative for predicting on-device focus, and that performance could be improved further by contextualising attentive behaviour using information on the visual scene (see \autoref{fig:aodfeature}).
A particularly interesting direction for future work is attention span prediction, i.e.\ the regression task of predicting the actual duration of attention on the mobile device and in the environment.
Preliminary experiments on our dataset (not shown here) suggested that this task is currently too challenging -- at least with the sensors and features used in this work.
It will be interesting to study this task in more detail in the future and to see which sensors and features will help to increase performance on this task above chance level.

\paragraph{Potential Applications}

Automatic forecasting of user attention opens up a range of exciting new applications that could have paradigm-changing impacts on our everyday interactions with mobile devices.
Predicted attention shifts to a mobile device 
could, for example, be used to reduce interaction delays.
The device could turn back on pro-actively and load the previous screen content for a smooth transition, or help users to reorient themselves on the device screen. 
However, attentive user interfaces are also faced with situations where predicted attention shifts to a mobile device should be prevented.
Especially within face-to-face conversations in the real world, user interfaces could help us to keep our focus by giving an alert to avoid unkind behaviour when there is a predicted shift to one's own mobile phone.
While driving, crossing a road, or walking down a busy street, it is also desirable for mobile device users to avoid attention shifts to the mobile device, to prevent potentially hazardous situations.
Attention shift prediction, for example combined with a detection of dangerous situations using an body-worn egocentric camera, could suppress on-device alerts or notifications to avoid such attention shifts.

For attention shifts to the environment, attention forecasting could be used to pro-actively support the users and automatically pause a video even before the attention drifts away, so that the user does not miss a second.
Similar to face-to-face conversations, predicted shifts to the environment could be prevented by attentive user interfaces during Skype meetings, so as to keep eye contact. 
Alternatively, if a user really wants to finish a task, the attentive user interface could help the user to keep their attention on the device by changing the content or style of content presentation.

If the primary attentional focus is predicted to be on the mobile device, previously missed messages or notifications could be shown to the user.
Moreover, the user interface could suggest the next task to be performed by the user.
Similar to avoiding attention shifts in dangerous situations, future user interfaces could break longer attentional focus spans when potential threats are detected via a scene camera.
The aforementioned prediction of attention span would further extend application opportunities by allowing for temporally more fine-grained and targeted adaptations.

\paragraph{Limitations and Future Work}

Despite these promising results, our work also has several limitations.
First, while we only considered visual triggers, attention shifts to the environment can also be triggered by auditory stimuli.
An interesting direction for future work is to analyse both visual and auditory information for predicting mobile attention allocation.
Second, we only considered prediction of temporal attention characteristics, namely timing of attention shifts and primary attentional focus.
Future mobile attentive user interfaces could also predict ``where'' user attention will shift~\cite{zhang2017deep}.
Third, while all our predictions were clearly above chance level, performance has to further increase to make attention forecasting practically useful.
To improve performance, additional sensors for heart rate, galvanic skin response (GSR) or brain activity could be used.
Given the rapid development in sensor technology, some of the wearables used may no longer be needed in the future, or they may be replaced by more sophisticated ones, providing even better features for attention forecasting.
Also, the method itself could be improved, for example, by using spatio-temporal CNN features extracted from each frame~\cite{tran2015learning} that demonstrated superior performance in a variety of computer vision tasks.
Particularly interesting are features extracted from intermediate layers, as for example used for vision-based~\cite{huang2018predicting,ma2016going} or wearable sensor-based~\cite{ordonez2016deep} activity recognition.
Fourth, the current hardware setup is rather bulky (head-mounted mobile eye tracker, multiple cameras, mobile phone, laptop backpack), which might have influenced participants' attentive behaviour.
Therefore, investigating in-the-wild studies with participants' awareness about the recording will be an interesting future project~\cite{nasiopoulos2015wearable,risko2011eyes}
Fully integrating the required cameras is an important direction for future work, but likely to be feasible given recent advances in fully embedded head-mounted eye tracking~\cite{tonsen17_imwut}.

\section{Conclusion}
\label{sec:conclusion}

In this work we explored \textit{attention forecasting} -- the task of predicting future allocation of users' overt visual attention during interactions with a handheld mobile device.
We proposed three prediction tasks with direct relevance for future mobile attentive user interfaces, as well as a first computational method to predict key characteristics of attentive behaviour from device-integrated and wearable sensors.
We evaluated our method on a novel 20-participant dataset and
demonstrated its effectiveness in predicting attention shifts between the mobile device and the environment, as well as the primary attentional focus on the mobile device.
Our results demonstrate not only the feasibility but also the significant challenge of attention forecasting, and point towards a new class of user interfaces that pro-actively support, guide or even optimise for users' ever-changing attentive behaviour.

\section{Acknowledgements}
We would like to thank all participants for their help with the data collection, as well as Preeti Dolakasharia, Nahid Akhtar and Muhammad Muaz Usmani for their help with the annotation.
This work was funded, in part, by the Cluster of Excellence on Multimodal Computing and Interaction (MMCI) at Saarland University, Germany, as well as by a JST CREST research grant under Grant No.:~JPMJCR14E1, Japan.

\balance{}

\bibliographystyle{SIGCHI-Reference-Format}
\bibliography{references}

\end{document}